\documentclass[a4paper, 12pt]{article}

\usepackage[dvips]{graphics}
\usepackage[dvips]{epsfig}
\usepackage{latexsym}
\usepackage{amssymb}
\usepackage{cite}

\def\b{\begin{eqnarray}}
\def\e{\end{eqnarray}}
\def\s{\scriptscriptstyle}
\def\1{\hskip1pt}
\def\2{\hskip2pt}
\def\4{\hskip4pt}
\def\5{\hskip5pt}
\def\={\5 = \5}
\def\k{\kappa}

\def\da{\dot{a}}

\def\nn{\nonumber}

\def\da{\dot{a}}

\def\dg{\dot{g}}
\def\ddg{\ddot{g}}

\renewcommand{\thefootnote}{\fnsymbol{footnote}}

\begin{document}
\begin{center}
{\huge\textbf{Standard Cosmology on a Self-Tuning Domain Wall \\}}
\vspace {15mm}
{\large \bf Conall Kennedy\footnote{e-mail: conall@maths.tcd.ie}}  \\
\setcounter{footnote}{6}
\vspace {3mm}
{\it School of Mathematics, Trinity College, University of Dublin, Ireland} \\
\vspace{10mm}
{\large \bf Emil M. Prodanov\footnote{e-mail: prodanov@maths.tcd.ie} \\}
\vspace {3mm}
{\it National Centre for Scientific Research "Demokritos", Athens, Greece}
\end{center}
\renewcommand{\thefootnote}{\fnsymbol{footnote}}
\setcounter{footnote}{0}
\vspace{10mm}

\begin{abstract}

We investigate the cosmology of (4+1)-dimensional gravity coupled to a scalar field and a {\em bulk} anisotropic fluid within the 
context of the single-brane Randall-Sundrum scenario. Assuming a separable metric, a static fifth radius and the scalar to depend only on the fifth direction, we find that the warp factor is given as in the papers of Kachru, Schulz and Silverstein [hep-th/0001206, hep-th/0002121] and that the cosmology on a self-tuning brane is standard. In particular, for a radiation-dominated brane the pressure in the fifth direction vanishes.
\end{abstract}

\normalsize

\newpage

\section{Introduction}

Theories with extra dimensions where our four-dimensional world is a 
hypersurface (three-brane) embedded in a higher-dimensional spacetime and at 
which gravity is localised have been the subject of intense scrutiny since 
the work of Randall and Sundrum \cite{rs}. The main motivation for such models 
comes from string theory where they are reminiscent of the Ho\v{r}ava-Witten 
solution \cite{hw} for the field theory limit of the strongly-coupled $E_8 
\times E_8$ \linebreak heterotic string. The Randall--Sundrum (RS) scenario 
may be modelled \cite{gubser,hollo} by coupling gravity to a 
scalar field and mapping to an equivalent supersymmetric quantum mechanics 
problem. A static metric is obtained with a warp factor determined by the 
superpotential. A generalisation to non-static metrics was considered by 
Bin\'{e}truy, Deffayet and Langlois (BDL) who modelled brane matter as a 
perfect fluid delta-function source in the five-dimensional 
Einstein equations \cite{bdl}. 
However, this resulted in non-standard cosmology in that the square of the 
Hubble parameter on the brane was not proportional to the density of the 
fluid. Other cosmological aspects of ``brane-worlds'' have been considered in 
\cite{cosmo}.\\
In this letter, we investigate RS-type single brane cosmological solutions of five-dimensional 
gravity coupled to a scalar field which we assume to depend only on the fifth dimension. 
We further assume that the fifth dimension is static and infinite in extent. We also include a {\it bulk} anisotropic fluid with energy-momentum 
tensor $\hat{T}^A_B(\rho) = \mbox{diag\1}(-\rho, p, p, p, P)$ 
and equations of state $P = \tilde{\omega} \rho, \2 p = \omega \rho$. Assuming a separable metric, we find that the warp factor is given as in the papers of Kachru, Schulz and Silverstein (KSS) \cite{kachru1,kachru2}. We also find that the cosmology on a self-tuning brane is standard but that the pressure in the fifth direction is constrained by the relation $\tilde{\omega} = \frac{3\omega-1}{2}$. In particular, we find that the pressure in the fifth direction vanishes for a radiation-dominated brane with $\omega=1/3$.

\section{The Model}
We consider a single, thin brane at $r=0$, as in KSS \cite{kachru1}. The action for the gravity and scalar part of the model is:
\b
S&=&S_{gravity}+S_{bulk} +S_{brane}\5 , \nonumber \\
S_{gravity} &=&
\frac{1}{2 \hat{\k}_{\s 5}^2} \int d^4 x  d r \2 \sqrt{-\hat{g}} \2 \hat{R} \5 ,
\nonumber \\
S_{bulk}&=&\int d^4 x d r \2   \sqrt{-\hat{g}} \2\left(-\frac{4}{3}\2\hat{g}^{AB}\partial_A\Phi\partial_B\Phi -U(\Phi)\right) \5 , \nonumber \\
S_{brane}&=& \int d^4 x\sqrt{-g^{\s (4)}}\2\left(-V(\Phi)\right) \5 ,
\e 
where $\hat{g}_{AB}$ is the five-dimensional metric, $g^{\s (4)}_{ij}$ is the induced metric on the brane and the tension of the brane is parametrised by $V(\Phi)$. 

We assume a separable metric with flat spatial three-sections on the brane:
\b
\label{metric}
ds^2 &=& \hat{g}_{AB}dy^Ady^B \nn \\&=&e^{2A(r)}\left(-dt^2 + g(t)\delta_{ab}dx^adx^b\right)+ dr^2 \5 .
\e
This is a natural generalisation of the 4$d$ flat Robertson-Walker metric to
 a RS context and is a special case of the BDL ansatz (see \cite{bdl}) with $n(t,r)=e^{A(r)}\2,\2a(t,r)=e^{A(r)}g^{1/2}(t)\2,\2 b(t,r)=1$ in conventional notation.

We shall also make the ansatz that both the potentials $U(\Phi)$ and $V(\Phi)$ are of Liouville
 type (see, for instance, \cite{reall}):
 \b
 \label{pot}
 U(\Phi) & = & U_0 e^{\alpha\Phi} \5, \nonumber \\
 V(\Phi) & = & V_0e^{\beta\Phi} \5,
 \e
 where $U_0$ and $V_0$ are constants.

 The stress-tensor for the scalar is 
 \b
 \hat{T}^A_B(\Phi) = \check{T}^A_B + \tilde{T}^A_B \5 ,
 \e
 where
 \b
 \check{T}^A_B= \frac{8}{3}\2\partial^A \Phi\partial_B\Phi  - \delta^A_B\Bigl(\frac{4}{3}\partial^C \Phi\partial_C\Phi + U(\Phi)\Bigr) \5 ,
 \e
 and
 \b
 \tilde{T}^A_B=-\frac{\sqrt{-g^{\s (4)}}}{\sqrt{-\hat{g}}}\2
  V(\Phi) \2 \delta(r) \2 g^{\s (4)}_{ij} \2 \delta^{iA} \2 \delta^j_B \5 ,
  \e
  where there is no sum over the indices $i$ and $j$. We shall assume that $\Phi$ depends only on $r$.

  The bulk fluid has the stress-tensor \cite{wes}:
  \b
  \hat{T}^A_B(\rho) = \mbox{diag\1}(-\rho, p, p, p, P)
  \e
  in the comoving coordinates $y^A$. $\rho$ is the density and $p$ and $P$ are the pressures in the three spatial directions on the brane and in
  fifth dimension, respectively. The anisotropy can be considered as a result of the mixing of two interacting perfect fluids \cite{oli}.

  \section{The Solutions}
  We now proceed to solve Einstein's equations $\hat{G}^A_B=\hat{\k}^2_{\s 5}(\hat{T}^A_B(\Phi)+\hat{T}^A_B(\rho))$ given the above ansatze. 

  If we take a linear combination of the $00$- and $11$-components of Einstein's equations then the following equation results:
  \b
  \label{1}
   \frac{\dg^2}{g^2} 
    -
  \frac{\ddg}{g} - \hat{\k}^2_{\s 5}\1
  e^{2A} (\rho + p) = 0 \5 .
  \e
  Therefore, we see that $\rho$ and $p$ must be of the form 
  \b
  \rho(t,r)&=&e^{-2A(r)}\left(\tilde{\rho}(t)+F(t,r)\right) \5 , \\
  p(t,r)&=&e^{-2A(r)}\left(\tilde{p}(t)-F(t,r)\right) \5 ,
  \e
  for arbitrary $F(t,r)$.
  However, it is normal to assume the equation of state $p=\omega\rho$, where $\omega$ is constant in the range $-1\leq \omega\leq 1$. In the generic case $\omega\neq -1$ this implies that $F$ should be zero. We shall assume this also to be so in the special case $\omega=-1$. Furthermore, we shall also assume $P=\tilde{\omega} \rho$. Equation (\ref{1}) then reduces to 
  \b
  \label{1mod}
  \frac{\dg^2}{g^2} 
   -
  \frac{\ddg}{g} -\hat{\k}^2_{\s 5} \1
  (1+\omega)\1\tilde{\rho} = 0 \5 .
  \e

  Given $F=0$, the $00$-component of Einstein's equations separates into
  \b
  \label{4}
  \frac{3}{4} \2 
  \frac{\dg^2}{g^2} - \hat{\k}^2_{\s 5}\2 \tilde{\rho} &=&  C \5, \\ \nn \\
  \label{5}
  (6A^{\prime 2} +3A^{\prime\prime}) + \frac{4\2\hat{\k}^2_{\s 5}}{3} \2
  \Phi^{\prime 2} + \hat{\k}^2_{\s 5}  \2 U  + \hat{\k}^2_{\s 5}
  \2 V \2 \delta(r) \1 &=& \1 C\2e^{-2A}\5 ,
  \e  
  where $C$ is the separation constant.

  The $rr$-equation also splits in two:
  \b
  \label{6}
 \frac{3}{2} \2 \frac{\ddg}{g}+ \hat{\k}^2_{\s 5}\2\tilde{\omega}\2\tilde{\rho} &=& D\2, \\ \nn \\
  \label{8}
  6A^{\prime 2} - \frac{4\2\hat{\k}^2_{\s 5}}{3}\2 \Phi^{\prime 2} +
  \hat{\k}^2_{\s 5}\2 U  &=& D\2 e^{-2A} \5 ,
  \e
  where $D$ is another separation constant.

Equation (\ref{8}) allows us to recast (\ref{5}) in the form
\b
\label{alt5}
3A'' +\frac{8\2\hat{\k}^2_{\s 5}}{3}\Phi^{\prime 2}+(D-C)\2e^{-2A}+\hat{\k}^2_{\s 5}
  \2 V \2 \delta(r) = 0 \5 .
\e
We shall see below that in fact $D=2\2C$.

 In addition, the equation of motion for the scalar field 
  \b
  \label{eqm}
  \frac{8}{3}\2\hat{\nabla}^2 \Phi - \frac{\partial U(\Phi)}{\partial \Phi} -
  \frac{\sqrt{-g^{\s (4)}}}{\sqrt{-\hat{g}}} \2 \frac{\partial V(\Phi)}
  {\partial \Phi} \2 \delta(r) = 0 \5 ,
  \e
  results in the equation
  \b
  \label{bulk}
  \frac{8}{3}\Phi^{\prime\prime}+\frac{32}{3}A^\prime \Phi^\prime - \alpha \2 U - \beta \2 V \delta(r)
  & = & 0\5 ,
  \e
  
Note that the scalar field equation of motion implies that $\hat{\nabla}^A\check{T}_{AB} =0$ (and, conversely, off the brane only). This, in turn, implies that the fluid equations of motion $\hat{\nabla}^A\hat{T}_{AB}(\rho) =0$ are automatically satisfied. 
  
\vskip8pt
\noindent{\bf \underline{The Warp Factor}} 
\vskip.3cm
\noindent Equations (\ref{8}), (\ref{alt5}) (with $D=2C$) and (\ref{bulk}) have been extensively studied in \cite{kachru1,kachru2,youm,kanti,csaki2}. The self-tuning domain wall (solution (I) of \cite{kachru1}) is given by 
\b
U=C=0 \5 , \5 \beta \neq \pm \frac{1}{a} \5 ,
\e 
\b
\label{scalar}
\Phi(r) &=& a\2\epsilon\log(d-cr) \5,  \\
\label{warp}
A(r)&=&\frac{1}{4}\log(d-cr)-e \5 ,
\e 
where $a=3/(4\2\sqrt{2}\2\hat{\k}_{\s 5})$ and $\epsilon$ is a sign that takes opposite values either side of the brane at $r=0$. The parameters $c$, $d$ and $e$ are constants of integration that can also differ either side of the brane. For (\ref{scalar}) to make sense, we require $d>0$. The continuity of $\Phi$ and $A$ across the brane requires
\b
d_+d_-= 1 \5 ,
\e
\b
e_+= \frac{1}{4}\log d_+ \5 , \5 e_-=\frac{1}{4}\log d_- \5 ,
\e
where we have chosen the convention $A(0)=0$ and denoted constants defined on the right (left) side of the the brane with a $+\1 (-)$ subscript. The jump condtions implied by (\ref{alt5}) and (\ref{bulk}) result in the relations
\b
c_+&=&-\frac{2}{3}\hat{\k}_{\s 5}^2 \2d_+\2(a\beta\epsilon_+ -1)\2V_0\2e^{a\beta\epsilon_+\log d_+} \5 ,\nonumber \\
c_-&=&-\frac{2}{3}\hat{\k}_{\s 5}^2 \2d_-\2(a\beta\epsilon_+ +1)\2V_0\2e^{a\beta\epsilon_+\log d_+} \5 .
\e
The solution is self-tuning because given $d_+$, $\epsilon_+ =\pm 1$ and $\beta \neq \pm 1/a$, there is a Poincar\'{e}-invariant four-dimensional domain wall for any value of the brane tension $V_0$; $V_0$ does not need to be fine-tuned to find a solution.

Other warp factors are possible both when $C=0$ and when $C\neq0$. Solution (II) of \cite{kachru1} with $U=0$ and solution (III) of the same reference with $U\neq 0$ are examples of the former case. The solution presented in \cite{kachru2} with $U=0$ provides an example the latter. 
  
\vskip8pt
\noindent{\bf \underline{The Cosmology}}
\vskip.3cm
\noindent Adding equations (\ref{4}) and (\ref{6}) gives:
\b
\label{c1}
\dg^2 + 2g\ddg + \frac{4}{3} \2 \hat{\k}^2_{\s 5} \2
g^2 (\tilde{\omega} -  1)\2 \tilde{\rho} = \frac{4}{3}\1(D+C)\1g^2 \5.
\e
On the other hand, using (\ref{4}) in (\ref{1mod}) we obtain:
\b
\label{c2}
\dg^2 + 2g\ddg + 2 \2 \hat{\k}^2_{\s 5} g^2 \2 (\omega-1)\2\tilde{\rho} = 4\1C\1g^2 \5.
\e
Since $\tilde{\rho}$ is generically a function of $t$ (rather than a constant) the above two equations imply the relations
\b
D&=&2\2C \\
\label{ttwiddle}
\omega &=&\frac{1}{3}\2(1+2\tilde{\omega}) \5 .
\e
Relation (\ref{ttwiddle}) previously appeared in \cite{us}. In particular, it implies that an isotropic (perfect) fluid ($P=p$) is stiff, that is, $\omega=\tilde{\omega}=1$. The attribute ``stiff'' refers to the fact that the velocity of sound in the fluid is equal to the velocity of light. It should be noted that the case of a bulk cosmological constant ($\omega=\tilde{\omega}=-1$) is not covered here; however, it corresponds to the choice $U(\Phi)=constant$ instead.

    Using (\ref{4}), equation (\ref{c1}) may be expressed alternatively as
    \b
\label{final}
     \tilde{\omega} \2 \dg^2 + 2g\2\ddg  = \frac{4}{3}\1(\tilde{\omega}+2)\1C\1g^2 \5 ,
     \e
     with the solutions\footnote{These solutions, in the particular case of an isotropic fluid, appeared in a different setting in \cite{keski}.}
\b
\label{expand}
 g \sim \left\{ \begin{array}{ll}\sinh^{2q}(\sqrt{\frac{C}{3q^2}}\2 t) \5 \5 \5 &C>0 \5,\\
t^{2q} \5 \5\5 &C=0 \5 , \\ 
\sin^{2q}(\sqrt{\frac{|C|}{3q^2}}\2 t) \5 \5 \5 &C<0 \5, \end{array}\right.
\e
where $q=1/(2+\tilde{\omega})=2/(3(1+\omega))=q_{standard}$.  

From (\ref{4}) we see that the density $\tilde{\rho}$ (which is actually the density of the fluid on the brane since we have defined $A(0)=0$) is positive: 
\b
\tilde{\rho}(t) = \left\{ \begin{array}{ll}\hat{\k}_{\s 5}^{-2}\2C\2\sinh^{-2}(\sqrt{\frac{C}{3q^2}}\2 t) \5 \5 \5 &C>0 \5,\\
\frac{3\2\hat{\k}^{-2}_{\s 5}\2 q^2}{t^2} \5 \5\5 &C=0 \5 , \\ 
\hat{\k}_{\s 5}^{-2}\2|C|\2\sin^{-2}(\sqrt{\frac{|C|}{3q^2}}\2 t) \5 \5 \5 &C<0 \5.\end{array} \right.
\e
When $C\geq0$, equation (\ref{final}) also allows the de-Sitter solutions $g=e^{\pm 2 \sqrt{C/3}\2 t}$. These solutions have vanishing density $\tilde{\rho}$ and were discussed in \cite{gubser,kachru1,kachru2,desitter}. For the case $C=0$, we obtain conventional cosmology $H=\da/a \propto \sqrt{\tilde{\rho}}$ on the brane with evolution at the standard rate. 

     Of particular note is the case of radiation-dominated fluid on the brane ($\omega=1/3$). From (\ref{ttwiddle}) we see that the pressure in the fifth direction vanishes and the stress tensor is then:
     \b
     \hat{T}^A_B(\rho) = e^{-2A(r)}\2\tilde{\rho}(t)\2\mbox{diag\1}(-1, \frac{1}{3},\frac{1}{3} ,\frac{1}{3} , 0) \5 ,
     \e
     with $q_{standard}=1/2$.

\section{Summary}
The self-tuning domain wall, with warp factor given by (\ref{warp}), has vanishing separation constant $C$ and therefore expands according to the power law (\ref{expand}) at the standard rate and exhibits conventional cosmology when coupled to a {\em bulk} anisotropic fluid. The pressure of the fluid in the fifth direction, $P$, is related to the isotropic pressure on the brane, $p$, via equation (\ref{ttwiddle}) and vanishes for a radiation-dominated brane.

\section*{Acknowledgements}

We are grateful to George Savvidy, Siddhartha Sen and Andy Wilkins for useful discussions and a critical reading 
of the manuscript. \\
E.M.P. is supported by EU grant HPRN-CT-1999-00161.


\begin{thebibliography}{99}

\bibitem{rs}
L. Randall and R. Sundrum:
\5 {\it A Large Mass Hierarchy from a Small Extra Dimension}.
Phys. Rev. Lett. {\bf 83}, 3370--3373 (1999), hep-ph/9905221; \newline
L. Randall and R. Sundrum:
\5 {\it An Alternative to Compactification}.
Phys. Rev. Lett. {\bf 83}, 4690--4693 (1999), hep-th/9906064.

\bibitem{hw}
P. Ho\v{r}ava and E. Witten:
\5 {\it Heterotic and Type I String Dynamics from Eleven Dimensions}.
Nucl. Phys. {\bf B460}, 506--524 (1996), hep-th/9510209; \newline
P. Ho\v{r}ava and E. Witten:
\4 {\it Eleven-Dimensional Supergravity on a Manifold with Boundary}.
Nucl. Phys. {\bf B475}, 94--114 (1996), hep-th/9603142.

\bibitem{gubser}
O. DeWolfe, D. Z. Freedman, S. S. Gubser and A. Karch:
\5 {\it Modelling the Fifth Dimension With Scalars and Gravity}.
hep-th/9909134.

\bibitem{hollo}
C. Csaki, J. Erlich, T. J. Hollowood and Yu. Shirman:
\5 {\it Universal Aspects of Gravity Localised on Thick Branes}.
hep-th/0001033.

\bibitem{bdl}
P. Bin\'{e}truy, C. Deffayet and D. Langlois:
\5 {\it Non-conventional Cosmology from a Brane-Universe}. hep-th/9905012.

\bibitem{cosmo}
N. Kaloper and A. Linde:
\5 {\it Inflation and Large Internal Dimensions}. Phys. Rev. {\bf D59}
101303 (1999), hep-th/9811141; \newline
J. M. Cline, C. Grojean and G. Servant:
\5 {\it Cosmological Expansion in the Presence of an Extra Dimension}.
Phys. Rev. Lett. {\bf 83}, 4245--4247 (1999), hep-ph/9906523; \newline
D. J. H. Chung and K. Freese:
\5 {\it Cosmological Challenges in Theories with Extra Dimensions and Remarks on the Horizon Problem}. Phys.  Rev. {\bf D61}, 023511 (2000), hep-ph/9906542; \newline
H. B. Kim and H. D. Kim:
\5 {\it Inflation and Gauge Hierarchy in Randall--Sundrum Compactification}.
Phys. Rev. {\bf D61} 064003 (2000), hep-th/9909053; \newline
P. Kanti, I. I. Kogan, K. A. Olive and M. Pospelov:
\5 {\it Cosmological 3-Brane Solutions}. Phys. Lett. {\bf B468}, 31--39 (1999),
hep-ph/9909481; \newline
E. E. Flanagan, S.-H. Henry Tye and I. Wasserman:
\5 {\it Cosmological Expansion in the Randall-Sundrum Brane World Scenario}.
hep-ph/9910498; \newline
U. Ellwanger: \5 {\it Cosmological Evolution in Compactified
Ho\v{r}ava--Witten Theory Induced by Matter on the Branes}. hep-th/0001126;
\newline
R. N. Mohapatra, A. P\'erez-Lorenzana and C. A. de S. Pires:
\5 {\it Inflation in Models with Large Extra Dimensions Driven by a Bulk
Scalar Field}. hep-ph/0003089; \newline
M. Br\"andle, A. Lukas and B. A. Ovrut:
\5 {\it Heterotic M-Theory Cosmology in Four and Five Dimensions}.
hep-th/0003256; \newline
B. Grinstein, D. R. Nolte and W. Skiba:
\5 {\it Adding Matter to Poincar\'{e}-Invariant Branes}. Phys. Rev. {\bf D62}, 086006 (2000), hep-th/0005001.

\bibitem{kachru1}
S. Kachru, M. Schulz and E. Silverstein:
\5 {\it Self-tuning Flat Domain Walls in 5d Gravity and String Theory}. Phys. Rev. {\bf D62} (2000) 045021, hep-th/0001206.

\bibitem{kachru2}
S. Kachru, M. Schulz and E. Silverstein:
\5 {\it Bounds on Curved Domain Walls in 5-D Gravity}. Phys. Rev. {\bf D62} 085003 (2000), hep-th/0002121.

\bibitem{reall}
H. A. Chamblin and H. S. Reall:
\5 {\it Dynamic Dilatonic Domain Walls}. Nucl. Phys. {\bf B562}, 133--157
(1999), hep-th/9903225.

\bibitem{wes}
H. Liu and P. S. Wesson:
\5 {\it Exact Solutions of General Relativity derived from 5-D Black-Hole Solutions of Kaluza-Klein Theory}. J. Math. Phys. {\bf 33} (1992) 3888.

\bibitem{oli}
S. R. Oliveira:
\5 {\it Model of Two Perfect Fluids for an Anisotropic and Homogeneous Universe}. Phys. Rev. {\bf D40} (1989) 3976.

\bibitem{youm}
D. Youm:
\5 {\it Bulk Fields in Dilatonic and Self-Tuning Flat Domain Walls}. Nucl. Phys. {\bf B589}, 315-336 (2000), hep-th/0002147.

\bibitem{kanti}
P. Kanti, K. A. Olive and M. Pospelov:
\5 {\it Static Solutions for Brane Models with a Bulk Scalar Field}. Phys. Lett. B481 (2000) 386, hep-ph/0002229.

\bibitem{csaki2}
C. Csaki, J. Erlich, C. Grojean and T. J. Hollowood:
\5 {\it General Properties of the Self-tuning Domain Wall Approach to the Cosmological Constant Problem}. Nucl. Phys. {\bf B584}, 359-386 (2000), hep-th/0004133.

\bibitem{us}
C. Kennedy and E. M. Prodanov:
\5 {\it Standard Cosmology from Sigma-Model}. Phys. Lett. {\bf B488} 11-16 (2000), hep-th/0003299.

\bibitem{keski}
K. Enqvist, E. Keski-Vakkuri and S. R\"{a}s\"{a}nen:
\5 {\it Constraints on Brane and Bulk Ideal Fluid in Randall-Sundrum Cosmologies}. hep-th/0007254.

\bibitem{desitter}
N. Kaloper:
\5 {\it Bent Domain Walls as Braneworlds}. Phys. Rev. {\bf D60} (1999) 123506, hep-th/9905210; \newline
T. Nihei:
\5 {\it Inflation in the Five-dimensional Universe With an Orbifold Extra
Dimension}. Phys. Lett. {\bf B465}, 81--85 (1999), hep-ph/9905487;\newline
M. Gremm:
\5 {\it Thick Domain Walls and Singular Spaces}. Phys. Rev. {\bf D62} (2000) 044017, hep-th/0002040;\newline
J. E. Kim and B. Kyae:
\5 {\it Exact Cosmological Solution and Modulus Stabilization in the Randall-Sundrum Model with Bulk Matter}. Phys. Lett. {\bf B486} (2000) 165, hep-th/0005139.








\end{thebibliography}
\end{document}